# COVID-19 and the stock market: evidence from Twitter

Rahul Goel[a], Lucas Javier Ford[b], Maksym Obrizan[b,1], Rajesh Sharma[a]

[a]Institute of Computer Science, University of Tartu, Tartu, Estonia
[b]Kyiv School of Economics, Kyiv, Ukraine

Abstract

COVID-19 has had a much larger impact on the financial markets compared to previous epidemics because the news information is transferred over the social networks at a speed of light. Using Twitter's API, we compiled a unique dataset with more than 26 million COVID-19 related Tweets collected from February 2nd until May 1st, 2020. We find that more frequent use of the word "stock" in daily Tweets is associated with a substantial decline in log returns of three key US indices – Dow Jones Industrial Average, S&P500, and NASDAQ. The results remain virtually unchanged in multiple robustness checks.

Declarations of interest: none

## 1. Introduction

COVID-19 led to substantial disruption on the financial markets, and stock markets in developed countries demonstrated the worst performance since 1987 (Zhang, Hu and Ji, 2020). The speed of access to new information is often quoted among key reasons why markets reacted much more forcefully compared to previous epidemics such as the 1918 Spanish Flu (Baker er al., 2020). Twitter, as a key supplier of news information in real time, has been actively used to study financial markets for the last 10 years (Bollen, Mao, and Zeng, 2011).

To study the effect of COVID-19 on financial markets, we collected 29,469,349 tweets from 7,875,334 unique users using Twitter's API from February 2nd until May 1st, 2020. Given the controversies in applying sentiment analysis (Azar and Lo, 2016; Sul, Dennis, and Yuan, 2017), we instead rely on a simple metric, namely, the share of the word "stock" mentions in daily tweets related to COVID-19. On the next step, we identified an autoregressive model for each of the three primary US indices: i) Dow Jones Industrial Average, ii) S&P500, and iii) NASDAQ. Finally, we expand the autoregressive models by adding an indicator variable capturing possible regime switching in the US financial markets and our measure of the "stock" word mentions.

We find that possible regime switching after the first reported US case of COVID-19 is significant at 1% only for the NASDAQ index. However, we observe a statistically significant negative association between the frequency of the word "stock" mentions and the performance of all three major US market indices, and this result remains robust in multiple alternative specifications.

In Section 2 we review previous literature on the effects of coronavirus and Twitter sentiment on the financial markets. Section 3 and 4 describe the dataset and methods. Section 5 presents our findings and the last section concludes.

---

[1]Corresponding author at: Kyiv School of Economics, 04050, Dmytrivska 92-94, Kyiv, Ukraine. *E-mail address:* mobrizan@kse.org.ua

## 2. Previous literature

In our review, we consider two main streams of the relevant literature: the effect of COVID-19 on financial markets and prior studies of tweets in relation to stock returns. Baker et al. (2020) compare the COVID-19 outbreak with the 1918 Spanish Flu using data from the S&P500 and Dow Jones Industrial Average (DJIA) Indices. They observe that the US stock market reacted much more forcefully to COVID-19 compared to previous epidemics. The factors that might explain such a difference include the speed of access to new information, the complexity of the networks that characterize modern economies, and the restrictive containment policies that governments have chosen as a response. Zhang, Hu and Ji (2020) notice that in March 2020, the US stock market hit the circuit breaker mechanism four times in ten days (which before that was only triggered once in 1997 since its introduction in 1987). However, unconventional policies, such as unlimited quantitative easing in the United States, can create long term problems.

Stock market predictions based on opinions from the micro-blogging social media Twitter have been capturing researchers' attention from various fields for about a decade. According to the Efficient Market Hypothesis news is the primary driver of stock market prices. Since new information is not predictable, stock market prices cannot be predicted with more than 50% accuracy because, in this case, prices follow a random walk. Contrary to this common view, Bollen, Mao, and Zeng (2011) used data from OpinionFinder and Google-Profile of Mood States (GPOMS) to determine that public mood states are predictive of the DJIA closing value with 87.6% accuracy. In a more recent study, Ruan, Durresi and Alfantoukh (2018) construct a user-to-user trust network and categorize users by the reputation and weight of their opinion. The authors find that their trust network approach could amplify correlations between Twitter's sentiment valence and the abnormal returns for the eight most mentioned firms in their data.

Ranco et al. (2015) find a significant relationship between Twitter sentiment and abnormal returns when volume peaks, even when these peaks are given for non-obvious events. Furthermore, the direction of DJIA abnormal returns is implied by the sentiment polarity of Twitter peaks. The authors argue that, while social media and market behavior is a worthy research field, the results depend heavily on the accurate selection of the data. Shen, Liu, and Zhang (2018) use Twitter's daily happiness sentiment by quintiles as a proxy for sentiment dynamics to study its relationship with the skewness of stock returns of 26 international stock market index returns. Their results suggest that the skewness of the highest-happiness quintile is significantly larger than that of the lowest-happiness quintile.

Despite a consensus among researchers on the power of measuring sentiment, some caveats need to be considered. For instance, Azar and Lo (2016) argue that serious questions can be posed about the quality of data taken from social media, where both informed and non-informed users can equally participate in discussisons about financial markets. In support of the previous argument, Sul, Dennis, and Yuan (2017) use a standard approach by weighing users' opinions on individual S&P500 firms by the number of followers. Surprisingly, opinions from users with a number of followers below the median that were not retweeted had the highest impact on predicting stock returns! Being intrigued by this last finding, we do not rely on the sentiment analysis of tweets in our analysis but rather on the relative frequency of mentions of the word "stock" in our sampled tweets.

## 3. Data description

In order to explore the effect of coronavirus (COVID-19) on the US stock market, we collected and stored a large sample of public tweets beginning February 1st, 2020, that matched a set of pre-defined search terms: "coronavirus", "coronaphobia", and "coronovirusoutbreak". Additional search terms were later added, such as "COVID-19" (On February 11th, 2020, COVID-19 was the official name given to novel coronavirus by WHO), "coronavirusitaly" (On February 21st, Italy COVID-19 outbreak) and "coronavirusitalla". The Twitter API was used to collect the dataset with additional language filter (tweets only in English were collected). The dataset spans more than three months (February 1st, 2020, to May 2nd, 2020) and covers 7,875,334 users. The dataset includes 29,469,349 tweets, of which 6,494,657 (or 22.0%) are original tweets, and the remaining are retweets.[2]

From those COVID-19 tweets, we have counted the number of times the word "stock" was mentioned each day as a share of total tweets related to the epidemic.[3] Our prior expectation is that if the word "stock" is mentioned more often on certain days, it may indicate some active developments in social sentiment, which may positively or negatively affect the stock market. Hence, we use the share of such mentions as our primary predictor.

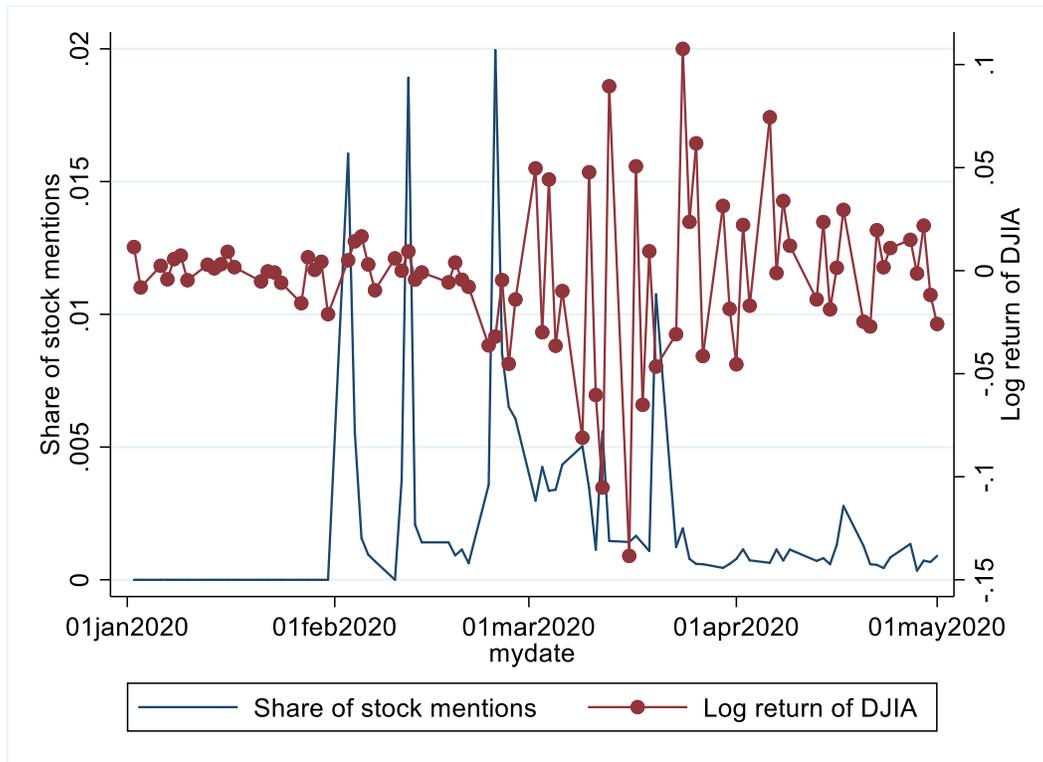

Figure 1. The relation between DJIA log returns and share of word "stock" mentions in tweets.

For our main dependent variables of interest, we downloaded three key US stock indices, namely Dow Jones Industrial Average, S&P500, and NASDAQ Composite for the last 10 years from Yahoo Finance.[4] Next, we converted the adjusted closing price for each index to continuous or log returns for analyses because returns have more desirable properties in financial time series analysis compared to stock prices (Tsay, 2005). We took 10 years of data for key stock indices in order to build a robust time series model of stock market behavior in the US. The data for our main predictor – the share of the "stock" word mentions in tweets related to COVID-19, is only available from February 1st, 2020 to May 2nd, 2020. Since COVID-19 only appeared recently, we converted missing values of our key predictor to 0. Figure 1 shows the connection between the DJIA index and our key predictor.

---

[2] Due to a technical problem tweets were not recorded on March 18th, 2020 so this day is removed from the sample.
[3] We have also experimented with using other keywords (such as "finance") and using counts of mentions instead of shares in all tweets. Based on the results of our trials we chose share of counts as the most robust measure of sentiment on Twitter.
[4] We used R package "quantmod" and the series are available from https://finance.yahoo.com/.

An attentive reader may raise a concern that our key variable only captures stock market uncertainty following the unfolding epidemic. In order to address this justified concern, we also include an indicator variable taking a value of 1 starting from January 20th, 2020 – the first confirmed COVID-19 case in the US and zero before that date. Table 1 provides descriptive statistics for key variables of interest.

Table 1. Descriptive statistics for key variables between May 3rd, 2010 and May 1st, 2020

| Variable | # | Mean | Std. Dev. | Min | Max |
| --- | --- | --- | --- | --- | --- |
| Log returns of DJIA | 2,516 | .0003259 | .0108602 | -.1384181 | .1076433 |
| Log returns of S&P500 | 2,516 | .0003615 | .0110008 | -.1276522 | .0896832 |
| Log of returns NASDAQ | 2,516 | .0005106 | .0122101 | -.1314915 | .0893470 |
| Share of "stock" word mentions in tweets | 2,516 | .0000700 | .0007772 | 0 | .0199528 |
| Date after 1st COVID-19 case in US | 2,516 | .0282194 | .165632 | 0 | 1 |

*Notes: Authors' calculations based on the final dataset.*

## 4. Methods

Autoregressive models play a prominent role in modeling financial time series such as stock and index returns, interest and exchange rates, as well as many macroeconomic series including GDP, unemployment, and inflation. Autoregressive models are built on the premise that past $p$ values of a financial time series jointly determine the conditional expectation of the time series $r_t$:

$$r_t = \varphi_0 + \varphi_1 r_{t-1} + \cdots + \varphi_p r_{t-p} + a_t, \qquad (1)$$

where $a_t$ is a white noise series. Partial autocorrelation function and Akaike Information Criterion (AIC) are two standard methods used to identify the order of the AR(p) process (Tsay, 2005).

In order to estimate the effect of tweets on the three key stock indices, we first build a standard autoregressive model for each index without tweets. Based on the partial autocorrelation functions, we select the 1st and 7th lag for Dow Jones Industrial Average and S&P500 and only the 1st lag for NASDAQ Composite Index. After identifying the order of autoregressive models, we add our key predictor – the share of mentions of "stock" word in the total number of tweets. It is possible that stock market behavior is driven by regime-switching rather than mentions on Twitter. To allow for such a possibility, we include an indicator variable taking a value of 1 after the date of the first COVID-19 case in the US.[5]

## 5. Results

Table 2 shows the results of the estimated model for three key stock indices in the USA. The results show that log returns on Dow Jones Industrial Average and S&P500 indices are not statistically different at 5% after the first reported US case of COVID-19. However, NASDAQ log returns are higher (significant at 1% level) after this date but the size of the effect is rather small (only 0.2% points). In addition, the first autoregressive term is negative and statistically significant for all three stocks, while the 7th lag is positive and significant when included indicating a strong mean reversion property of the financial markets.

---

[5] https://www.nejm.org/doi/full/10.1056/NEJMoa2001191

Table 2. Model of daily log returns on DJIA, S&P500, and NASDAQ indices between May 3rd, 2010 and May 1st, 2020

| Variables | DJIA | S&P500 | NASDAQ |
|---|---|---|---|
| Share of "stock" word mentions in tweets | -1.268*** | -1.310*** | -1.264*** |
| | (-8.533) | (-8.517) | (-7.504) |
| Date after 1st COVID-19 case in US | 0.000 | 0.001* | 0.002*** |
| | (0.640) | (1.666) | (2.826) |
| Constant | 0.000 | 0.000* | 0.001** |
| | (1.600) | (1.732) | (2.188) |
| AR(1) | -0.148*** | -0.151*** | -0.127*** |
| | (-20.968) | (-19.454) | (-13.910) |
| AR(7) | 0.121*** | 0.104*** | |
| | (14.556) | (10.798) | |
| Observations | 2516 | 2516 | 2516 |

*Notes: t-statistics in parentheses. * p<0.10, ** p<0.05, *** p<0.01.*

The most interesting is that the share of "stock" word mentions in tweets related to COVID-19 has a statistically significant negative effect at 1% on all three stock market indices. What is more, is that the magnitude of the effects is very similar across three different measures of stock market behavior. In particular, one standard deviation increase in the share of the "stock" word mentions in tweets is associated with 9.1% of standard deviation increase in DJIA log returns, 9.3% standard deviation increase in S&P500 log returns, and 8.0% standard deviation increase in NASDAQ log returns.

In the Online Appendix, we report the results of three robustness checks to verify that results are stable. First of all, we added indicators for trading days from Monday to Thursday to capture possible variations in trading strategies on various days. Next, we limited the sample only to the last 5 and 1 years. In all cases, coefficients are remarkably stable and always statistically significant at 1%.[6]

**6. Discussion and conclusions**

To the best of our knowledge, this is the first paper that identifies a new relevant predictor of stock market behavior in the US during the uncertain times of COVID-19 epidemics. Using 1% of publicly available tweets between February 1st, 2020, and May 2nd, 2020, we have identified those related to COVID-19 epidemics. On the next step, we computed the number of times when the word "stock" was used in coronavirus tweets during the day and divided this number by the total number of COVID-19 tweets sampled on that day. Then we identified the AR models for three key US stock indices – Dow Jones Industrial Average, S&P500, and NASDAQ Composite Index. Finally, we estimated the autoregressive models with the share of "stock" word mentions and an indicator variable from the first COVID-19 case in the US to capture possible regime switching. We found that the time indicator for the beginning of the coronavirus epidemic in the US is not significant at 1% for 2 out of 3 stock indices. However, we found a highly significant negative effect of the "stock" word mentions in COVID-19 tweets, which is also robust in many alternative specifications.

---

[6] When we limit the sample to the days when mentions of the "stock" word are available we still get negative coefficients but they are no longer statistically significant, probably because of a small sample size of just 62 observations.

**CRediT authorship contribution statement**

**Rahul Goel:** Data curation, Software, Writing - original draft. **Lucas Javier Ford:** Writing - original draft. **Maksym Obrizan:** Formal analysis, Writing - original draft. **Rajesh Sharma:** Conceptualization, Writing - original draft.

**Acknowledgments**

Rahul Goel and Rajesh Sharma acknowledge funding for this research by ERDF via the IT Academy Research Programme and SoBigData++.

**Online Appendix**

Table A1. Model of daily log returns on DJIA, S&P500, and NASDAQ indices between May 3rd, 2010 and May 1st, 2020 with indicators of weekdays

| Variables | DJIA | S&P500 | NASDAQ |
| --- | --- | --- | --- |
| Share of "stock" word mentions in tweets | -1.271*** | -1.313*** | -1.271*** |
|  | (-8.535) | (-8.525) | (-7.526) |
| Date after 1st COVID-19 case in US | 0.000 | 0.001* | 0.002*** |

|                          | (0.636)     | (1.651)     | (2.815)     |
| ------------------------ | ----------- | ----------- | ----------- |
| Indicator for Monday     | -0.001      | -0.001      | -0.000      |
|                          | (-0.771)    | (-0.915)    | (-0.479)    |
| Indicator for Tuesday    | 0.001       | 0.001       | 0.001       |
|                          | (0.807)     | (0.892)     | (1.198)     |
| Indicator for Wednesday  | 0.000       | 0.000       | 0.001       |
|                          | (0.360)     | (0.398)     | (0.779)     |
| Indicator for Thursday   | 0.000       | 0.000       | 0.000       |
|                          | (0.092)     | (0.047)     | (0.125)     |
| Constant                 | 0.000       | 0.000       | 0.000       |
|                          | (0.612)     | (0.666)     | (0.472)     |
| AR(1)                    | -0.148***   | -0.151***   | -0.127***   |
|                          | (-20.256)   | (-18.845)   | (-13.558)   |
| AR(7)                    | 0.121***    | 0.105***    |             |
|                          | (14.353)    | (10.728)    |             |
| Observations             | 2516        | 2516        | 2516        |

*Notes: t-statistics in parentheses. * p<0.10, ** p<0.05, *** p<0.01.*

Table A2. Model of daily log returns on DJIA, S&P500, and NASDAQ indices for the last 5 years

| Variables                                | DJIA        | S&P500      | NASDAQ      |
| ---------------------------------------- | ----------- | ----------- | ----------- |
| Share of "stock" word mentions in tweets | -1.174***   | -1.196***   | -1.289***   |
|                                          | (-6.123)    | (-6.638)    | (-6.927)    |
| Date after 1st COVID-19 case in US       | 0.000       | 0.001       | 0.002**     |
|                                          | (0.027)     | (0.813)     | (2.434)     |
| Constant                                 | 0.000       | 0.000       | 0.001       |
|                                          | (0.924)     | (0.940)     | (1.398)     |
| AR(1)                                    | -0.185***   | -0.201***   | -0.203***   |
|                                          | (-19.940)   | (-21.206)   | (-17.656)   |
| AR(7)                                    | 0.181***    | 0.185***    |             |
|                                          | (17.676)    | (16.010)    |             |
| Observations                             | 1259        | 1259        | 1259        |

*Notes: t-statistics in parentheses. * p<0.10, ** p<0.05, *** p<0.01.*

Table A3. Model of daily log returns on DJIA, S&P500, and NASDAQ indices for the last 1 year

| Variables                                | DJIA        | S&P500      | NASDAQ      |
| ---------------------------------------- | ----------- | ----------- | ----------- |
| Share of "stock" word mentions in tweets | -1.114**    | -1.131***   | -1.331***   |
|                                          | (-2.210)    | (-2.617)    | (-3.645)    |
| Date after 1st COVID-19 case in US       | -0.000      | 0.000       | 0.002       |
|                                          | (-0.094)    | (0.033)     | (0.748)     |
| Constant                                 | 0.001       | 0.001       | 0.001       |
|                                          | (0.172)     | (0.253)     | (0.422)     |
| AR(1)                                    | -0.283***   | -0.312***   | -0.403***   |
|  | (-10.204) | (-11.460) | (-13.651) |
| AR(7) | 0.251*** | 0.266*** |  |
|  | (9.458) | (9.574) |  |
| Observations | 254 | 254 | 254 |

*Notes: t-statistics in parentheses. * p<0.10, ** p<0.05, *** p<0.01.*